\documentclass[11pt]{article}
\usepackage{graphicx}
\usepackage{amsmath}
\usepackage{amssymb}
\usepackage{amsxtra}

\usepackage{subcaption, csquotes}

\usepackage[
	style=phys,
	biblabel=brackets,
	eprint=true,
	bibencoding=auto,
	backend=biber,
	sorting=none,
	maxbibnames=3,
	language=auto,
	autolang=other,
	doi=false,
	url=false
	]{biblatex}

\AtEveryBibitem{
	\clearfield{day}
	\clearfield{month}
	\clearfield{endday}
	\clearfield{endmonth}
}

\usepackage[colorlinks]{hyperref}

\title{Domain walls and strings formation in the early Universe}
\author{
    A. A. Kirillov \\
    AAKirillov@mephi.ru \\
    B. S. Murygin \\
    MuryginBS@gmail.com \\
    National Research Nuclear University MEPhI \\
    (Moscow Engineering Physics Institute)
    }

\date{}

\begin{document}
\maketitle

\begin{abstract}
    Solitons formation through classical dynamics of two scalar fields with the potential having a saddle point and one minimum in (2+1)-space-time is discussed. We show that under certain conditions in the early Universe both domain walls and strings can be formed even if scalar fields are inflaton ones.
    %Moreover, we show that only one minimum is needed for the formation of solitons.
\end{abstract}


% optionally

\section{Introduction}

Multi-field inflation models such as the hybrid inflation \cite{1994PhRvD..49..748L} or the natural inflation \cite{2005JCAP...01..005K, 2015JCAP...06..040P} may contain potentials of non-trivial forms. If potential has at least one saddle point, the field dynamics in such models may lead to formation of topologically non-trivial structures named solitons \cite{rajaraman, vilenkin, manton}. Moreover, under certain conditions, they may produce primordial black holes in the radiation era due to collapse of domain walls \cite{2019EPJC...79..246B} or loops of cosmic strings \cite{2019PhRvD..99j4028H} that affects the early Universe \cite{2018JCAP...11..008V}.

Previously, it was shown solitons may be formed in (1+1) space-time even potential has only one minimum and at least one saddle point \cite{2017JPhCS.934a2046G, 2018JCAP...04..042G}. In this paper, we continue study of the possibility in (2+1) space-time.

% Under certain conditions, these topological nontrivial structures might lead to primordial black holes production in the early Universe due to collapse of domain walls \cite{2001JETP...92..921R, 2005APh....23..265K} or loops of cosmic strings \cite{2019PhRvD..99j4028H}. Moreover, the solitons may dramatically affect on the early Universe \cite{2018JCAP...11..008V, 2019EPJC...79..246B}.%

%However, mechanism that we discuss in this paper applicable not only for inflaton fields, but for any two real scalar fields with sufficiently complex potential.

%Solitons are scalar field topological defect that might appear during classical evolution e.g. see \cite{1998hep.ph....7343K , 2000hep.ph....5271R} where formation of domain wall in potential with several minima is discussed. For more detailed consideration see e.g. \cite{vilenkin, manton, rajaraman}. Also such topological defects could form Primordial Black Holes \cite{2020arXiv200212778C}. In this paper we consider soliton production in potential with only one minima.

\section{Model in (2+1) space-time}

%We study classical evolution of system of two real scalar fields numerically. Also, in our calculations dependancy on $z$ axis coordinate is neglected, in other words we study evolution in (2+1) space-time. See also \cite{2018JCAP...04..042G, 2017JPhCS.934a2046G} where evolution of similar system for (1+1) case is studied.

Let us consider the dynamics of two real scalar fields $\varphi$ and $\chi$ with the Lagrangian of the system 
\begin{equation}
    \label{Lagrangian}
    \mathcal{L} = \frac{1}{2}g^{\mu\nu}
        \big(\partial_{\mu}\varphi\partial_{\nu}\varphi +
        \partial_{\mu}\chi\partial_{\nu}\chi\big) -
        \mathcal{V}(\varphi,\chi),
\end{equation}
where $g^{\mu\nu}$ is the Friedman-Robertson-Walker metric tensor with the cosmic scale factor $a(t)$. Then, the classical motion equations for $\varphi$ and $\chi$ in (2+1) space-time take the form
\begin{equation}
    \label{Motion}
    \begin{aligned}
        \varphi_{tt}-3H\varphi_t-\varphi_{xx}-\varphi_{yy} &= -\frac{\partial\mathcal{V}}{\partial{\varphi}},
        \\
        \chi_{{t}{t}}-3H\chi_t-\chi_{xx}-\chi_{yy} &= -\frac{\partial\mathcal{V}}{\partial\chi}.
     \end{aligned}
\end{equation}
Here, $H=\dot{a}/a$ is the Hubble parameter which is $H_{I} \sim 10^{13}$~GeV during the inflation and becomes smaller in the radiation era. For equations~\eqref{Motion}, the Hubble parameter plays a role of a friction term, and its time dependence does not affect our conclusions. Thus, we assume it remains constant after the end of the inflation. In addition, the Hubble parameter $H$ gives a natural scale for all units. Therefore, we express all dimension variables in $H_I$ units.
 
To solve the system \eqref{Motion}, we have to define initial and boundary conditions. We choose the initial conditions in the form
\begin{equation}
    \label{InitialConditions}
    \begin{aligned}
        \varphi(x,y,0) &=\mathcal{R}\cos{\Theta}+\varphi_1, & \varphi_t(x,y,0) &= 0; 
        \\
        \chi(x,y,0) &=\mathcal{R}\sin{\Theta}+\chi_1, & \chi_t(x,y,0) &= 0,
    \end{aligned}
\end{equation}
where 
\begin{equation}
    \label{InitialConditionsLegend}
    \mathcal{R}(r) = \mathcal{R}_0 \cosh^{-1}{\cfrac{r_0}{r}},
    \quad
    \Theta = \theta.
\end{equation}
It sets correspondence between the fields space $(\varphi,\chi)$ and the physical plane $(x,y)$. Here, the point $(\varphi_1, \chi_1)$ corresponds to the center of the initial fields area in the form of the circular disk with the radius $\mathcal{R}(r)$ and the polar angle $0\leq\Theta\leq2\pi$, $r=\sqrt{{x}^2+{y}^2}$ and $\theta$ are a distance from the coordinate origin and a polar angle in a physical $xy$-plane, respectively, and  $\mathcal{R}_0$ and $r_0$ are positive parameters.

The boundary conditions are chosen as
\begin{equation}
    \label{BoundaryConditions}
    \begin{aligned}
        \varphi_x(\pm\infty,y,t) & = 0, & \varphi_y(x,\pm\infty,t)& = 0;
        \\
        \chi_x(\pm\infty,y,t) & = 0, & \chi_y(x,\pm\infty,t)& = 0.
    \end{aligned}
\end{equation}
 
We study classical evolution of the scalar fields $\varphi$ and $\chi$ with the potential used in \cite{2017JPhCS.934a2046G, 2018JCAP...04..042G}: 
\begin{equation}
  \label{Potential1}
  \mathcal{V}  = d(\varphi^2+\chi^2)+a\exp{\big[-b(\varphi-\varphi_0)^2-c(\chi-\chi_0)^2\big]},
\end{equation}
where $a$, $b$, $c$, $d$ are positive parameters. The parameter $a$ sets a height of a local maximum, $b$ and $c$ set its shape, and $d$ is responsible for a slope of the potential. The described potential has only one saddle point and one minimum, but could be easy modified to obtain any number of saddle points by adding terms like the last one.

Additionally, we consider the well-known potential ``tilted Mexican hat'' \cite{2015JCAP...06..040P}
\begin{equation}
  \label{Potential2}
  \mathcal{V}  = \lambda\biggl(\varphi^2+\chi^2-\frac{g^2}{2}\biggr)^2+\Lambda^4\Biggl(1-\frac{\varphi}{\sqrt{\varphi^2+\chi^2}}\Biggr),
\end{equation}
where $\lambda$, $g$, $\Lambda$ are positive parameters. The parameter $g$ sets a position of a circle of degenerate minima in the case of the Mexican hat without a tilt, $\lambda$ sets a height of a local maximum at the point $(\varphi_0,\chi_0)=(0,0)$ and $\Lambda$ sets a tilt of the potential. Note, a potential slope makes minima non-degenerate. However, non-degeneracy is not a necessary condition for solitons production.

The energy density of the system is given by
\begin{equation}
  \label{EnergyDensity}
  \rho = \frac{1}{2} \sum_{i} \big((\partial_i\varphi)^2 + (\partial_i\chi)^2\big) + \mathcal{V}(\varphi,\chi),
\end{equation}
where partial derivatives are taken over the variables $\{t, x, y\}$.
 
\section{Results}

\begin{figure}[!t]
    \centering
    \begin{subfigure}{0.47\textwidth}
      \includegraphics[width=\textwidth]{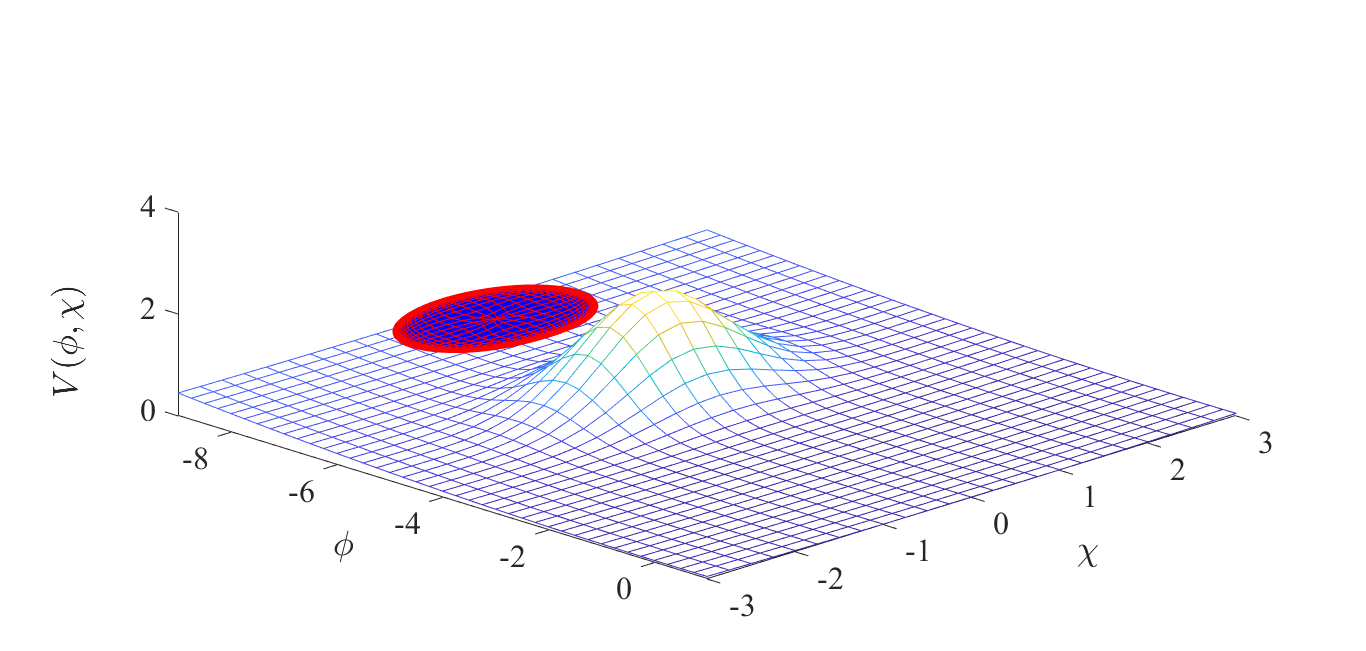} 
      \caption{The initial conditions for domain wall production.}
      \label{fig:V_in_dw}
    \end{subfigure}
    \hfil
    \begin{subfigure}{0.47\textwidth}
      \includegraphics[width=\textwidth]{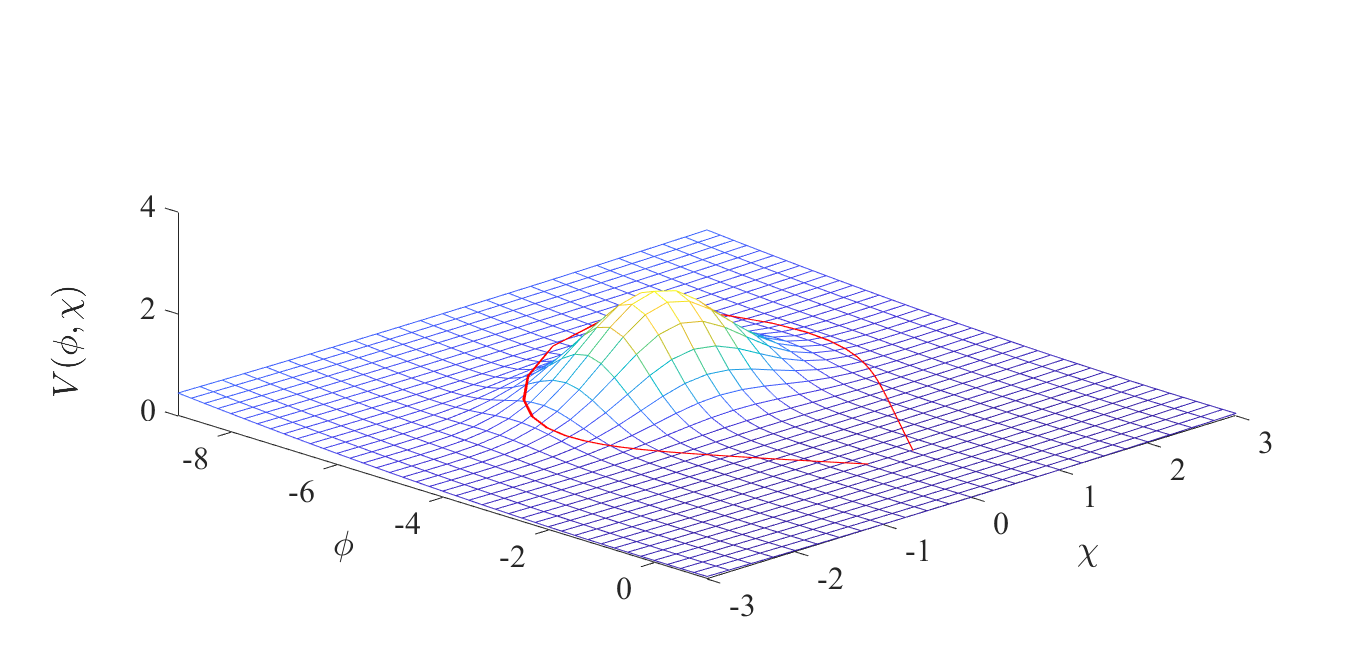}
      \caption{The final state showing the domain wall production.}
      \label{fig:V_fin_dw}
    \end{subfigure}
    \begin{subfigure}{0.47\textwidth}
      \includegraphics[width=\textwidth]{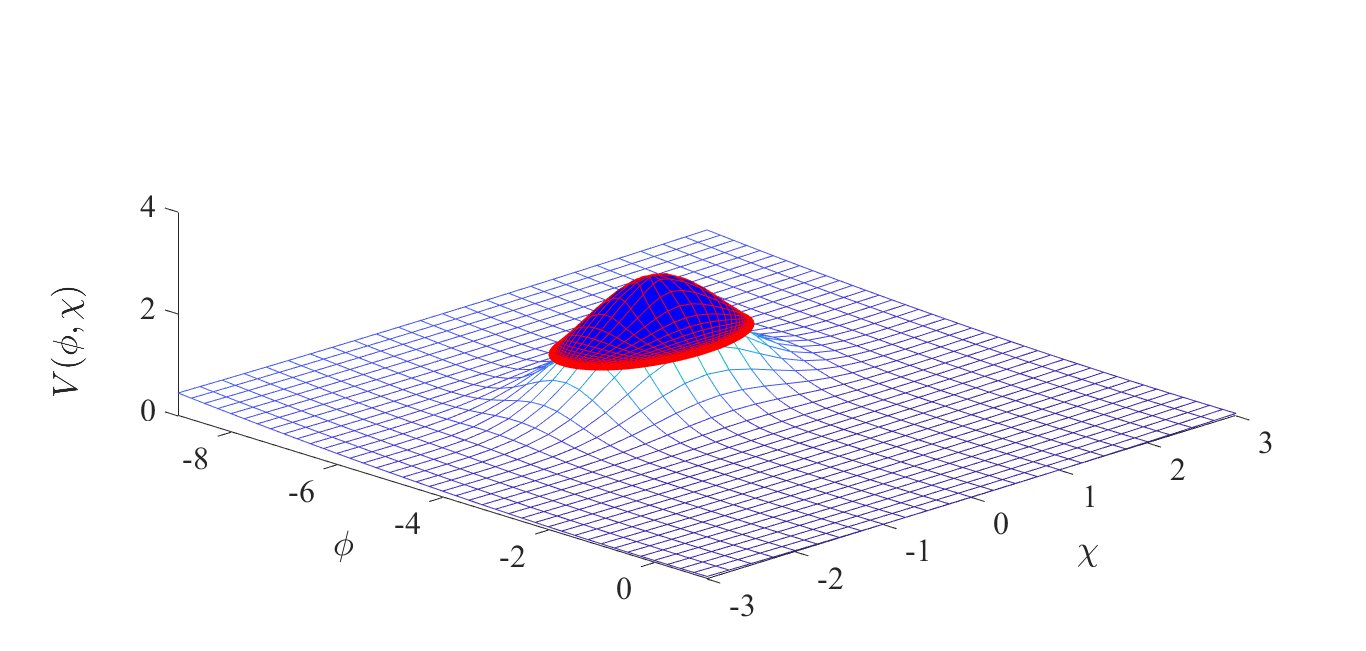}
      \caption{The initial conditions for string production.}
      \label{fig:V_in_str}
    \end{subfigure}
    \hfil
    \begin{subfigure}{0.47\textwidth}
      \includegraphics[width=\textwidth]{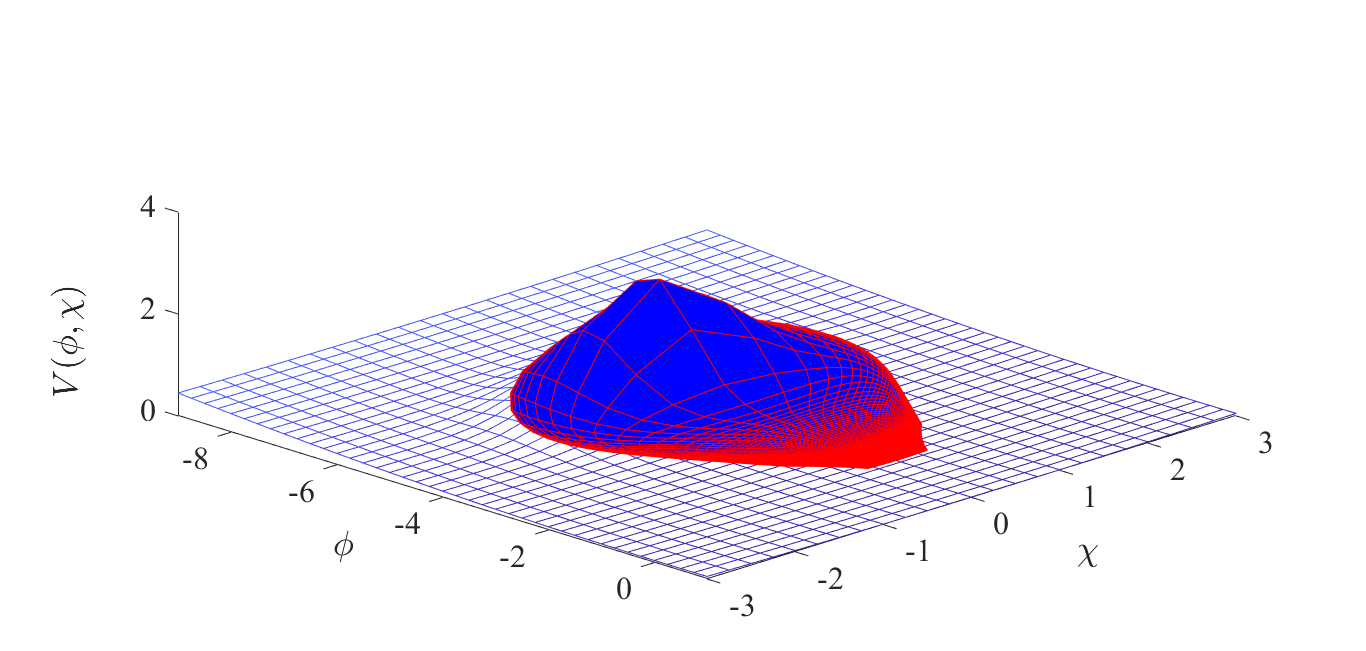}
      \caption{The final state showing the string production.}
      \label{fig:V_fin_str}
    \end{subfigure}
    \caption{The initial and final states of the fields configuration with the potential \eqref{Potential1} are shown.}
\end{figure}

For the potential \eqref{Potential1}, we 
%considered two cases differs only in initial location of the field configuration. 
choose the parameters as follows 
% In both cases potential parameters and all parameters except $\varphi_1$ in initial condition stay the same and set as follows 
$d=0.005$, $a=2$, $b=1$, $c=1$, $\varphi_0=-5$, $\chi_0=0$ and the parameters of the initial conditions \eqref{InitialConditions} $\mathcal{R}_0=1$, ${r}_0=1$, $\varphi_1 = -8$ and $\chi_1=0$ (all dimensional values are taken in $H_I$ units). The initial fields configuration is separated from the minimum by the peak of the potential, see fig.~\ref{fig:V_in_dw}. Note, the potential has the minimum at the point $(\varphi_\text{min}, \chi_\text{min})=(0,0)$. 

The final state of the fields evolution is presented in fig.~\ref{fig:V_fin_dw}. Due to the slope of the potential, the fields tend to reach its minimum; however, there are two possible ways to achieve it. Thus, the fields may bypass the peak from both sides. Note, we do not consider the situation when energy of the system is large enough to overcome the peak without bypassing. As a result of our calculations, the configuration reaches the minimum in both possible ways and stops in the equilibrium state.

\begin{figure}[!t]
    \centering
    \begin{subfigure}{0.47\textwidth}
      \includegraphics[width=\textwidth]{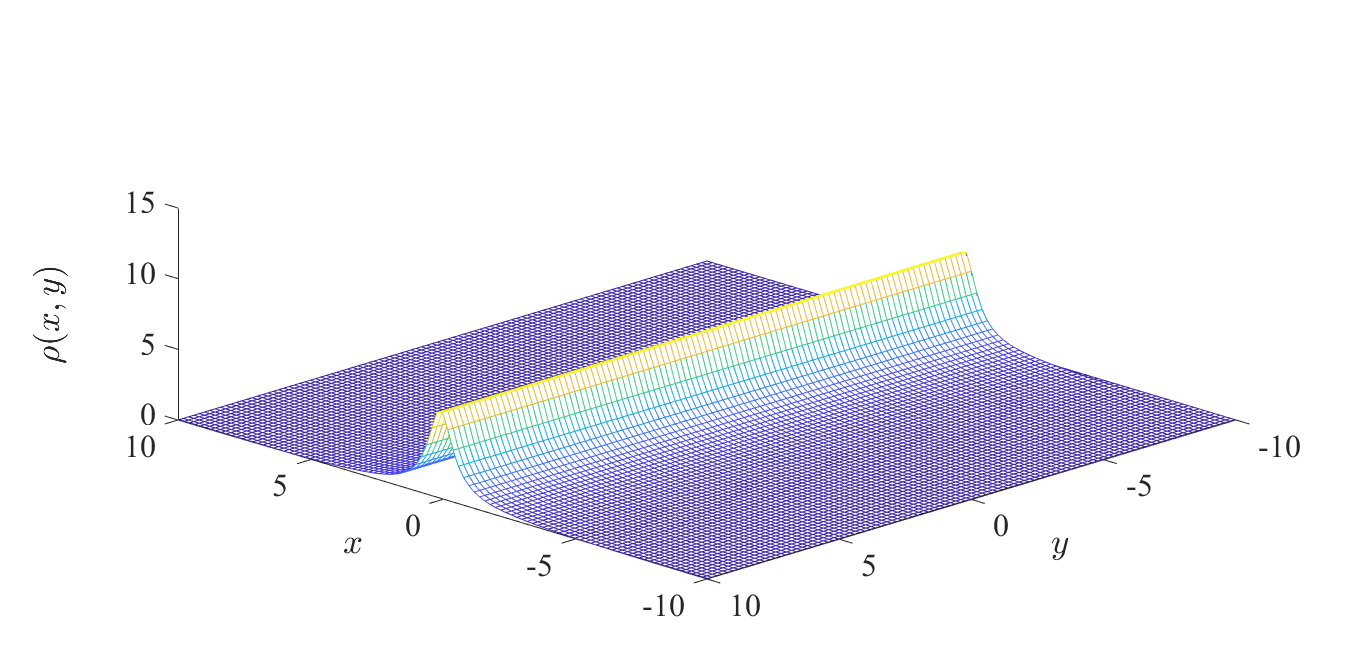} 
      \caption{The energy density of the domain wall.}
     \label{fig:rho_fin_dw}
    \end{subfigure}
    \hfil
    \begin{subfigure}{0.47\textwidth}
      \includegraphics[width=\textwidth]{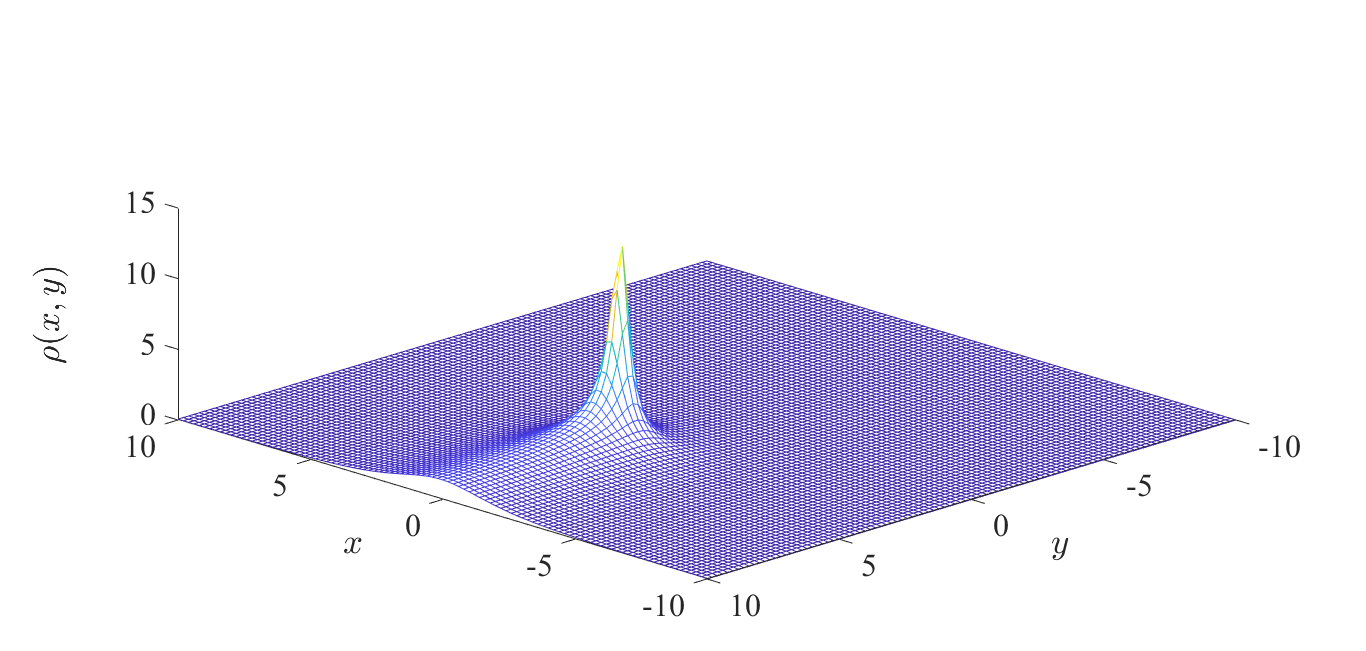}
      \caption{The energy of the string with the ridge.}
      \label{fig:rho_fin_str}
    \end{subfigure}
    \caption{The energy densities of the final states are shown.}
    \label{energy}
\end{figure}

%The determination of the type of soliton was carried out by comparing the obtained energy density distribution shown in fig.\ref{energy} with the similar assumed distribution for the corresponding type of soliton, i.e. fig.\ref{fig:rho_fin_dw} is a domain wall distribution and fig.\ref{fig:rho_fin_str} is a string energy density distribution. 
The energy density of the fields configuration determined by \eqref{EnergyDensity} is presented in fig.~\ref{fig:rho_fin_dw}. It corresponds to well-known type of solitons named domain walls and confirms the results of \cite{2018JCAP...04..042G} for the winding number $N=1$. This stable soliton may play a significant role in the early Universe. Closed domain walls with high enough energy could collapse due to surface tension and thus produce primordial black holes \cite{2001JETP...92..921R, 2005APh....23..265K}.


If we fix all parameters except $\varphi_1$, the other solitons type is obtained. We choose new value $\varphi_1=-5$, it changes the initial conditions in fields space $(\varphi,\chi)$, see fig.~\ref{fig:V_in_str}. 
%evolution goes from configuration shown in fig.\ref{fig:V_in_str} to the one shown in fig.\ref{fig:V_fin_str}. 
One can see, the initial state is now at the top of the peak. % and its only way to reach minima is to overcome the peak. 
The system evolution leads that the fields tend to the potential minimum. Thus, the configuration leaves the peak in all directions. The final state is shown in fig.~\ref{fig:V_fin_str}. The energy density of the formed soliton is presented in fig.~\ref{fig:rho_fin_str}, it corresponds to the formation of the string with the ridge. The last essentially distinguishes this soliton from well-known strings.

It is interesting to note that changing only one parameter $\varphi_1$ leads to the other soliton type. Thus, both domain walls and strings could be formed in one model. However, it does not take place for all possible sets of the parameters. Domain wall stability requires the second term of the potential \eqref{Potential1} corresponding to the local maximum height to be much bigger than the first one corresponding to the slope of the potential. It imposes boundaries on the parameters $a$ and $d$ showing whether domain walls may be formed.
%parameter that responsible for slope of the potential (in our case at least $10^3$ times bigger). However, it is only necessary for domain wall to be stable, otherwise it will decay at some instant depending on difference between two parameters mentioned earlier. 
In the case of strings formation, the parameter $\mathcal{R}_0$ may be restricted because the big value gives the system too large initial potential energy which may lead to destruction of solitons.
%However, strings may produce the primordial black holes due to loops collapse \cite{1996PhRvD..53.3002C, 2019PhRvD..99j4028H}. It should be kept in mind to avoid solitons overproduction in the early Universe if the fields system describes the inflation.  % And for the string case it always decay at some instant, however, this moment can be moved arbitrarily far depending on the values of the parameters.

\begin{figure}[!t]
    \centering
    \begin{subfigure}{0.47\textwidth}
      \includegraphics[width=\textwidth]{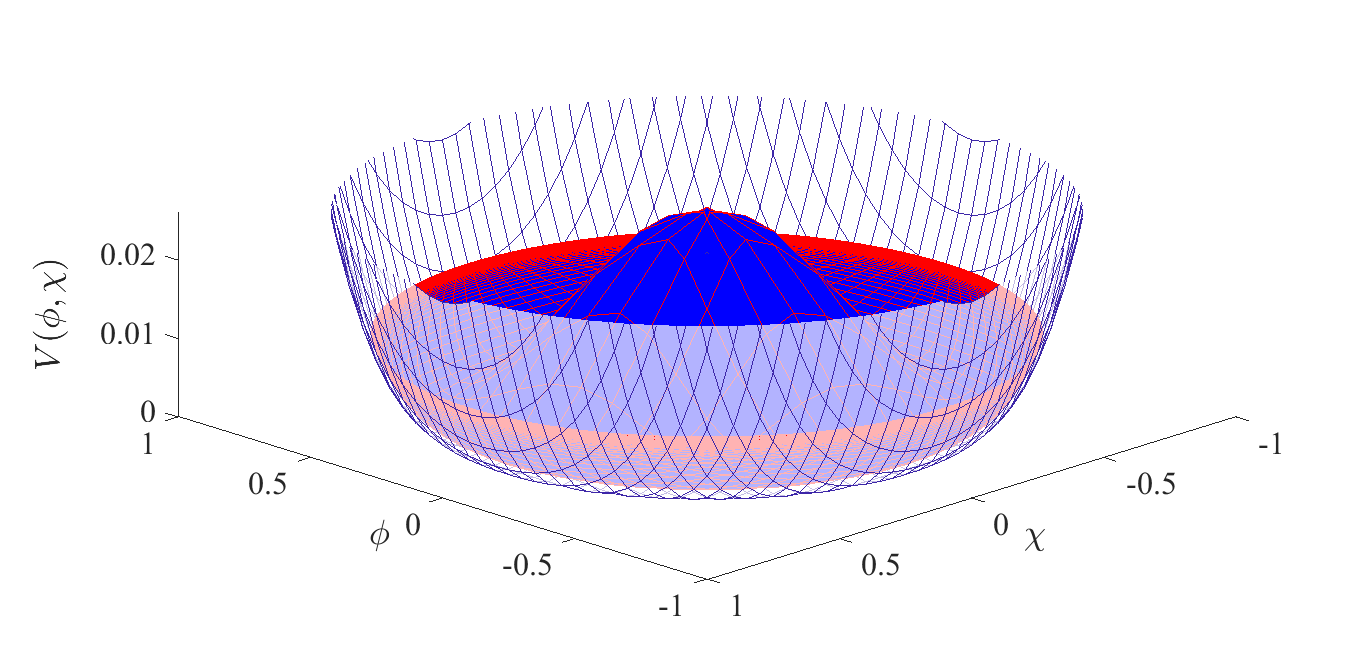}
      \caption{The initial fields distribution.}
      \label{fig:mexin}
    \end{subfigure}
    \hfil
    \begin{subfigure}{0.47\textwidth}
      \includegraphics[width=\textwidth]{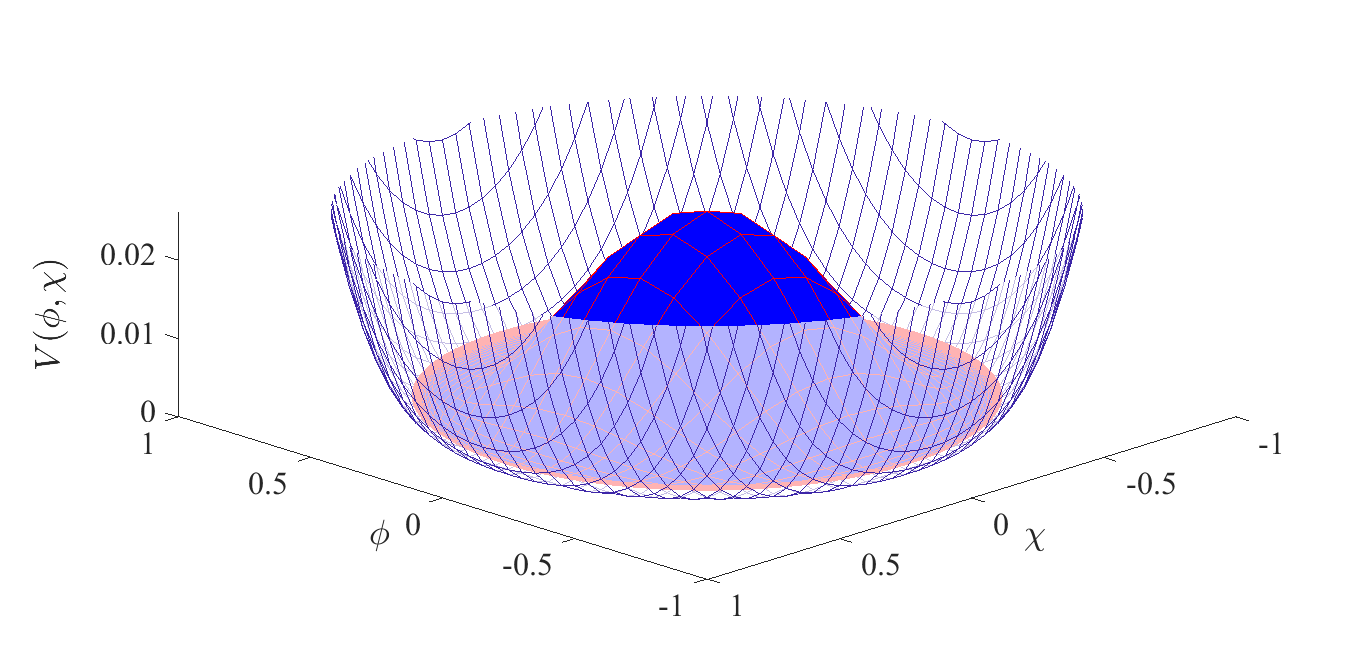}
      \caption{The final fields state.}
      \label{fig:mexfin}
    \end{subfigure}
    \caption{The initial and final states of the fields with the tilted Mexican hat potential \eqref{Potential2} are shown.}
    \label{evolution3}
\end{figure}

\begin{figure}[!t]
    \begin{minipage}[b]{0.48\textwidth}
        \includegraphics[width=\textwidth]{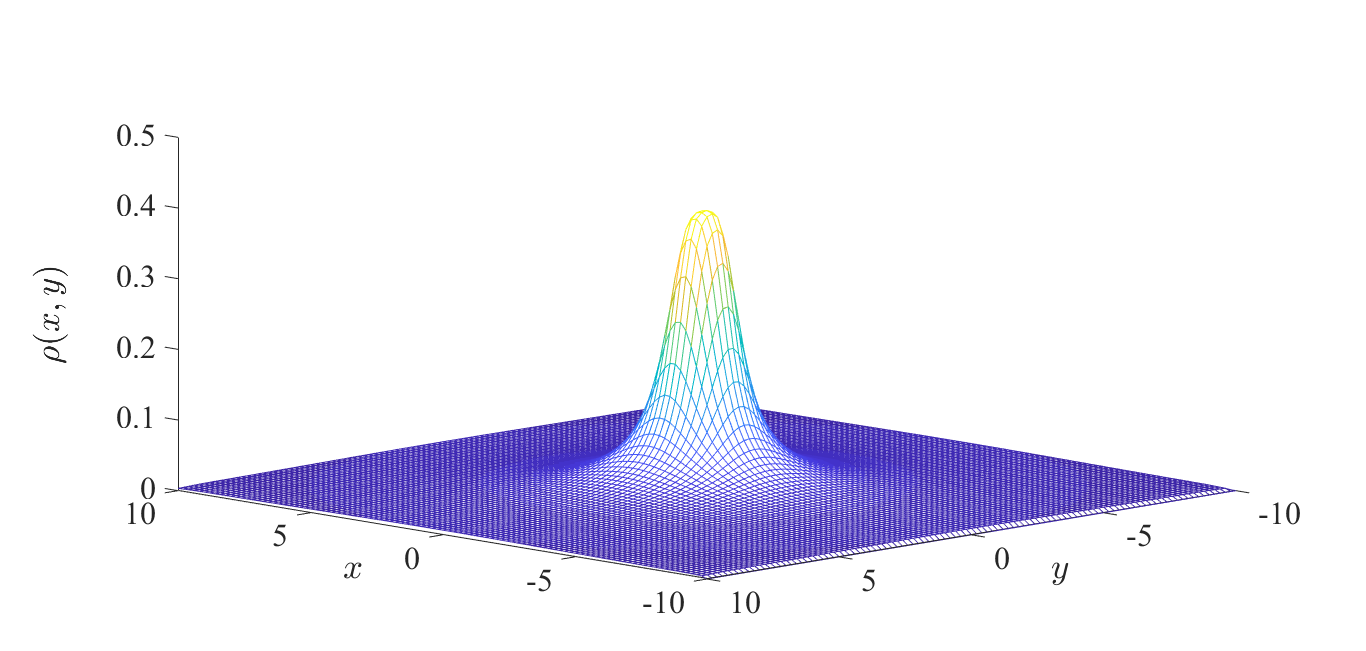}
    \end{minipage}
    \begin{minipage}[b]{0.48\textwidth}
        \caption{The final energy density distribution corresponding to string for tilted Mexican hat is shown.}
        \label{energy2}
    \end{minipage}
\end{figure}

%As mentioned before we study two different potentials. 
Finally, let us demonstrate solitons formation in the well-known tilted Mexican hat model \eqref{Potential2} describing the inflation. The possibility of domain walls production in this potential was considered in \cite{2018JCAP...04..042G}. Here we focus on the other solitons type. We choose the parameters of the potential as follows $g=1$, $\lambda=0.1$, $\Lambda=5\cdot10^{-13}$ and the parameters of the initial conditions $\mathcal{R}_0=0.9$, $r_0=1$, $\varphi_1=\chi_1=0$ (all in $H_I$ units). It is shown in fig.~\ref{fig:mexin} where the initial state of the fields is located on the peak top. Because the tilt is very small, fields configuration shrinks around the peak due to surface tension and stops when potential slope compensates it (see fig.~\ref{fig:mexfin}). The energy density of the final distribution is presented in fig.~\ref{energy2}. One can see, it corresponds to the string formation. The ridge is not seen due to the extremely small slope of the potential. If the parameter $\Lambda$ increases, the ridge appears. 


\section{Conclusion}

The solitons formation in the system of the scalar fields with potentials having a saddle point and one minimum was discussed. It is shown both domain walls and strings may be formed in the same model depending on the initial fields configuration. Note, the initial conditions may affect the solitons production even if the scalar fields are inflaton ones (it takes place if potential has at least one saddle point). In this case, it is important to check whether solitons appear in the model in order to avoid their overproduction in the early Universe, and, consequently, primordial black holes that may contradict observational data \cite{2020arXiv200212778C}.
%Conditions required to form solitons (especially for string formation) due to mechanism considered in this paper might be performed in many cases. Therefore this could provide useful method for limiting parameters in the inflation models. Along with solitons formations itself there is some motivation to study subsequent PBH formation from the solitons topological defects to check possible consequences of field dynamics of inflaton-like fields on further evolution of the Universe. However this a topic for a separate study.

\section*{Acknowledgement}

The authors are grateful to K.~M.~Belotsky, V.~A.~Gani, and S.~G.~Rubin for useful discussions. The work was supported by the grant RFBR No~19-02-00930.

\sloppy
\printbibliography

\end{document}